\documentclass[aps,preprint]{revtex4}
\usepackage[dvips]{graphicx}
\usepackage{amsmath}
\begin{document}
\draft

\title{\bf Missing level statistics and the power spectrum of level fluctuations analysis of three-dimensional chaotic microwave cavities}

\author{Micha{\l} {\L}awniczak, Ma{\l}gorzata Bia{\l}ous, Vitalii Yunko, Szymon Bauch, and Leszek Sirko}

\address{Institute of Physics, Polish Academy of Sciences, Aleja  Lotnik\'{o}w 32/46, 02-668 Warszawa, Poland}

\date{\today}

\bigskip

\begin{abstract}

We present an experimental study of missing level statistics of three-dimensional chaotic microwave cavities. The investigation is reinforced by the power spectrum of level fluctuations analysis which also takes into account the missing levels. On the basis of our data sets we demonstrate that the power spectrum  of
level fluctuations in combination with short- and long-range spectral fluctuations provides a powerful tool for the determination of the fraction of randomly missing levels in systems that display wave chaos such as the three-dimensional chaotic microwave cavities.  The experimental results are in good agreement with the analytical expressions that explicitly take into account the fraction of observed levels $\varphi$.  We also show that in the case of incomplete spectra with many unresolved states the above procedures may fail. In such a case the random matrix theory calculations can be useful for the determination of missing levels.
\end{abstract}

\bigskip
\maketitle
\section{Introduction}

Despite of extensive experimental studies of chaos in low-dimensional microwave systems such as one-dimensional (1D) microwave networks ~\cite{Hul2004,Lawniczak2010,Hul2012} and two-dimensional (2D) microwave cavities ~\cite{Stoeckmann90,Sridhar,Richter,So95,Stoffregen1995,Sirko1997,Hemmady2005,Dietz2015}, simulating quantum graphs and 2D billiards, respectively,  the studies devoted to three-dimensional (3D) chaotic microwave cavities (billiards) \cite{Sirko1995,Alt1997,Dorr1998,Dembowski2002,Tymoshchuk2007,Savytskyy2008} are very scarce due to experimental difficulties.
The microwave networks simulate the spectral properties of quantum graphs~\cite{Kottos1997,Kottos1999,Pakonski2001}, networks of 1D wires joined at vertices, owing to a direct analogy between the Telegraph equation describing a microwave network and the Schr\"odinger equation of the corresponding quantum graph \cite{Hul2004,Sirko2016}. They provide an extremely rich system for the experimental studies of the properties of quantum systems, that exhibit  chaotic dynamics in the classical limit. In the experiments with the 2D microwave billiards the analogy between the scalar Helmholtz equation and the Schr\"odinger equation of the corresponding quantum billiard is exploited.

 In the case of 3D microwave cavities there is no direct analogy between the vectorial Helmholtz equation and the Schr\"odinger equation, therefore,  in general, 3D microwave cavities cannot simulate 3D quantum systems. Nevertheless, experiments with 3D microwave cavities are very attractive and valuable because, as demonstrated in  an experimental study  by Deus {\it et al.} \cite{Sirko1995}, the distribution of eigenfrequencies of the 3D chaotic (irregular) microwave cavity displays behavior characteristic for classically chaotic quantum systems, viz., the Wigner distribution of the nearest neighbor spacings. Therefore, 3D chaotic microwave cavities appeared to be extremely useful in investigation of the properties of wave chaos.
  It is worth to point out that the perturbed cubic and rectangular cavities were studied in the microwave domain in Ref. \cite{Schroeder1954} while the  spectral statistics of acoustic resonances in 3D aluminum and quartz blocks were investigated in important papers by Weaver \cite{Weaver1989} and Ellegaard {\it et al.} \cite{Ellegaard1996}, respectively.  The systems investigated in Refs. \cite{Weaver1989,Ellegaard1996}  were characterized by high quality factors $Q\simeq 10^4-10^5$  therefore no missing modes were reported. In billiard systems, higher quality factors  were obtained in the experiments with superconducting microwave cavities \cite{Alt1997} while  in
normal conducting resonators the quality factors are much lower ($Q\simeq 10^3$) and the loss of some modes is either very likely or even inevitable \cite{Stoeckmann90}.
 In other experiments with chaotic 3D microwave cavities  \cite{Alt1997,Dorr1998,Dembowski2002} the periodic orbits, the distributions of the frequency shifts caused by the external perturbation and a trace formula have been studied. Properties of  nodal domains in a chaotic 3D microwave rough billiard were studied in Refs. \cite{Tymoshchuk2007,Savytskyy2008}. In the presence of strong absorption the experimental distribution $P(R)$ of the reflection
coefficient $R$ for the 3D microwave rough cavity was measured by {\L}awniczak {\it et al.}  \cite{Lawniczak2009} while  the
impedance and scattering variance ratios were evaluated by Yeh {\it et al.} \cite{Yeh2013} in the 3D microwave reverberating cavity. In both cases the experimental results appeared to be in good agreement with the Random Matrix Theory (RMT) predictions.
The three-dimensional chaotic cavities as well as the properties of random electromagnetic vector fields have been scarcely studied
theoretically \cite{Primack2000,Prosen1997,Arnaut2006,Gros2014}.

Due to small amplitudes of some resonances and the large density of states in 3D systems the loss of levels in an experiment is inevitable even in the regime of low absorption.
However, neither experimental nor theoretical studies of chaotic 3D microwave cavities which explicitly take into account missing levels have been reported so far.

A comparison of the spectral properties of quantum systems with the results of Gaussian Orthogonal Ensemble (GOE) or Gaussian Unitary Ensemble (GUE) in the RMT requires complete sequences of eigenvalues belonging to the same symmetry class~\cite{Bohigas1983,Bohigas1984}. Therefore, experimental determination of the chaoticity of a system on the basis of spectral fluctuation properties is in general far from simple \cite{Sieber1993,Dietz2014}.

 The situation has been significantly improved when a new procedure to obtain information on the chaoticity and time symmetry of classical systems from the spectral properties of the corresponding quantum systems in the presence of missing levels was developed and experimentally demonstrated for the microwave networks with broken \cite{Bialous2016} and preserved \cite{Dietz2017} time reversal symmetries. In the real physical systems such as  nuclei and molecules~\cite{Liou1972,Zimmermann1988,Frisch2014,Mur2015}, one always deals with the incomplete spectra, therefore, such a procedure is indispensable for their analysis~\cite{Enders2000,Agvaanluvsan2003,Bohigas2004}.  The impact of missing levels on the spectral fluctuation properties is particularly large for the long-range spectral fluctuations ~\cite{Bohigas2004}. Moreover, it was demonstrated numerically in Ref.~\cite{Molina2007} and confirmed experimentally \cite{Bialous2016,Dietz2017} that the power spectrum  of
level fluctuations ~\cite{Relano2002,Faleiro2004} is a powerful statistical measure, extremely sensitive to missing levels and the time symmetry of the system. Accordingly, in order to identify the fraction of missing levels and the symmetry of the system, we consider all these statistical measures.

\section{Experimental setup}

In the experimental investigations of the missing-level statistics and the power spectrum   of
level fluctuations we used an irregular 3D microwave cavity which was made of polished aluminum
type EN 5754, the density of which is 2.67 g/cm$^3$. It consists of a rough half-circle element (marked by (1) in Fig.~1(b)) of maximum height $h$ = 60 mm closed by
 two flat elements, the side ((2) in Fig.~1(b)) and the upper ones.  The third element at the bottom is a slightly convex inclined plate ((3) in Fig.~1(b)), the purpose of which is to remove  the bouncing balls orbits between the upper and bottom walls of the cavity (see Fig.~1 for the dimensions and shape
details). The rough element is described on the cross-section
plane by the radius function $R(\theta) = R_0 +\sum^M _{m=2} a_m
\sin(m\theta+\Phi_m)$, where the mean radius $R_0$ =10.0 cm, $M$ =
20, $a_m$ and $\Phi_m$ are uniformly distributed on [0.084, 0.091]
cm and [0, 2$\pi$], respectively, and 0$\leq\theta<\pi$.
To make different realizations of the 3D cavity an aluminium scatterer  was
inserted inside the cavity (see Fig.~1). The scatterer was mounted on the metallic axle that allowed for its controllable rotation.
The orientation of the scatterer inside the cavity was changed by turning the axle around in
$18$ equal steps, each having the value $\pi /9$ radians.

\begin{figure}[h!]
\includegraphics[width=1.0\linewidth]{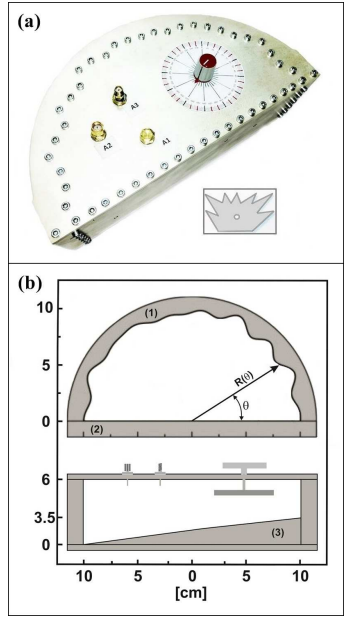}
\caption{
Panel (a): A photograph of the 3D microwave cavity showing the antenna positions A$_1$, A$_2$ and A$_3$.  For the measurement of the two-port scattering matrix $\hat S$ the vector network analyzer was coupled to the cavity through the  HP 85133-616 and HP 85133-617 flexible microwave cables via the two microwave antennas.  The inset shows the scatterer which was used to make  different realizations of the cavity. Panel (b) shows a sketch of the cavity on the cross-section plane (see the detailed description in the text) and the position of the scatterer inside the cavity.}
\label{Fig1}
\end{figure}

 The resonance frequencies of the cavity were obtained from the elements $S_{11}$, $S_{22}$, and $S_{12}$ of the two-port scattering matrix \cite{Lawniczak2015}

\begin{equation}
\label{Smatrix}
\hat S=\left[
\begin{array}{c c}
S_{11} & S_{12}\\
S_{21} & S_{22}
\end{array}
\right]
\end{equation}
 measured by an Agilent E8364B microwave vector network analyzer (VNA). The VNA was coupled to the cavity through the  HP 85133-616 and HP 85133-617 flexible microwave cables via the two microwave antennas. The three holes A$_1$, A$_2$ and A$_3$ in the upper plate of the cavity enable to perform measurements of the two-port scattering matrix $\hat S$ of the cavity in the frequency range $6-11$ GHz for the three different combinations of positions of the antennas.   During the measurements the unused antenna holes were plugged by the brass plugs. The diameter of the wire of the antennas was $0.9$~mm and they penetrated $6$~mm into the cavity.

 The distribution of the reflection coefficient $P(R)$ \cite{Lawniczak2009} revealed that the total absorption of the measured system is dominated by the internal absorption of the cavity, which is roughly 3 times larger than the absorption introduced by  a single antenna.

In Fig.~2 we show examples of the measured modules of the elements $|S_{11}|$ and $|S_{22}|$ of the two-port scattering matrix $\hat S$ (ports $A_1$ and $A_3$) of the 3D microwave cavity in the frequency range $7-8$ GHz  and $10-11$ GHz, respectively.  An inspection of the scattering matrix elements $|S_{11}|$ and $|S_{22}|$ clearly show a strong dependence of the recorded resonances on the positions of the antennas.  Some resonances at given antennas positions are very weak  or  not recorded at all. Therefore, due to too small amplitudes of some resonances and additionally strong, cubic dependence of the number of levels on the frequency $\nu$ (Weyl formula (\ref{Bloch})), which for larger $\nu$ may cause overlapping of some resonances, the missing levels in the measurements are unavoidable.

\begin{figure}[h!]
\includegraphics[width=1.0\linewidth]{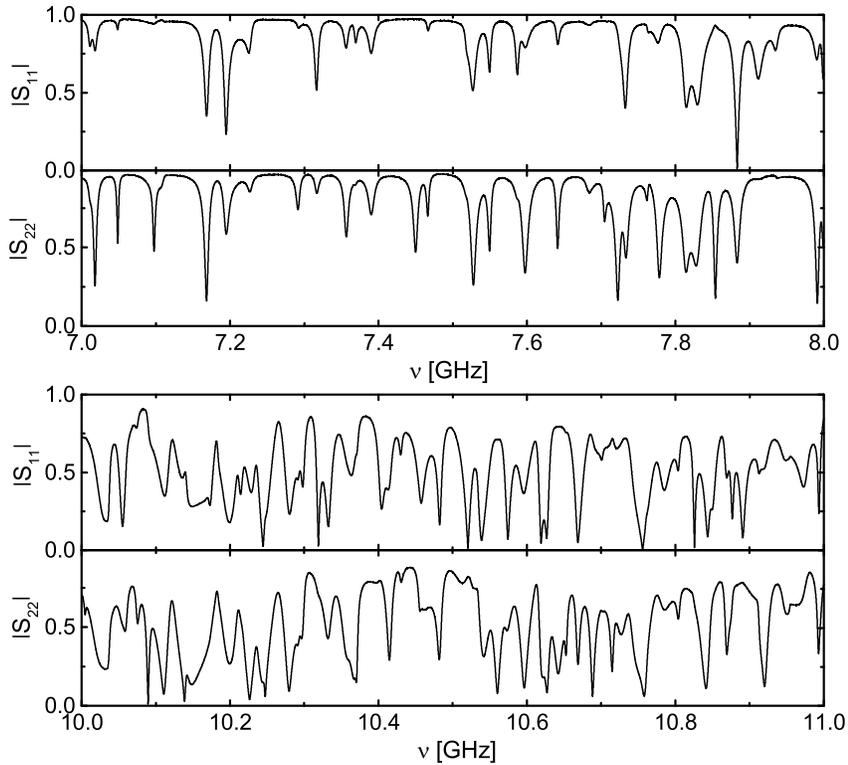}
\caption{
The measured modules of the elements $|S_{11}|$ and $|S_{22}|$ of the two-port scattering matrix $\hat S$ (ports $A_1$ and $A_3$) of the 3D microwave cavity in the frequency range $7-8$ GHz  and $10-11$ GHz, respectively.
}
\label{Fig2}
\end{figure}

\section{Fluctuations in the experimental spectra}

In order to perform an analysis of the spectral properties of the 3D microwave cavity the resonance frequencies $\nu_i$ need to be rescaled to eliminate system specific properties such as the volume and the surface of the cavity.  This is done with the help of the Weyl formula \cite{Balian,Balian1971,Baltes1972,Gros2014}

\begin{equation}
N(\nu)=A\nu^3-B\nu + C.
\label{Bloch}
\end{equation}
 An analysis of the electromagnetic modes in a 3D cavity with perfectly conducting walls showed that the transverse nature of the field and the boundary conditions lead to a cancelation of the surface term ($\sim \nu^2$), which is not present in this formula \cite{Balian1971}. In the formula~(\ref{Bloch}) only the coefficient $A=\frac{8}{3}\frac{\pi}{c^3}V$ can be easily calculated on the basis of the geometrical size of the cavity. Here, $V = (7.267\pm 0.012)\times 10^{-4}$ m$^3$ is the volume of the empty cavity reduced by the volume of the scatterer and  $c$ is the speed of light in vacuum. The geometric dependent coefficient $B$ depends on the surface curvature, local internal angles and the edges length of the cavity \cite{Gros2014}.  The constant $C$ also depends on the shape of the cavity.  For the simple case of a cube-shaped cavity it takes the value $C=\frac{1}{2}$ \cite{Baltes1972}.

 In general, it is difficult to determine the parameters $B$ and  $C$ analytically. Therefore, one may apply two possible approaches for evaluating the cumulative number of levels $N(\nu)$ for 3D irregularly shaped cavities. In the first one, applicable also when the geometrical size of the cavity is unknown or in the presence of missing levels, all required parameters $A$, $B$, and $C$ are obtained from a fit of the $N(\nu)$ to the experimental data. The second method can be used for complete spectra when the coefficient $A$ is known. In that case the parameters $B$ and $C$  are obtained from a fit of the $N(\nu)$ to the data. Consequently, because the cubic term in the Weyl formula ~(\ref{Bloch}) is dominant, e.g., the ratio   $A\nu^3/B\nu $ is approximately equal to 8 and 13 at the frequencies  7 GHz and 9 GHz, respectively, this method provides higher accuracy of calculations, especially when applied to the spectra with small number of resonances. Therefore, it was used to evaluate the total number of eigenvalues of the 3D cavity in the frequency range 6-11 GHz on the basis of nine complete spectra of the cavity measured in the narrower frequency range 7-9 GHz (see Section V). This allowed us  to estimate the fraction $\varphi$ of the observed levels independently from the missing level analysis.

 Accordingly, in the unfolding procedure the rescaled eigenvalues  $\epsilon_i$ are determined from the resonance frequencies $\nu_i$ as $\epsilon_i=N\left(\nu_i\right)$.

Commonly used measures for short-range spectral fluctuations are the nearest-neighbor spacing distribution (NNSD), that is, the distribution of the spacings between adjacent eigenvalues, $s_i=\epsilon_{i+1}-\epsilon_i$ and the integrated nearest-neighbor spacing distribution $I(s)$. For long-range spectral fluctuations we will consider the spectral rigidity of the spectrum $\Delta_3(L)$, given by the least-squares deviation of the integrated resonance density of the eigenvalues from the straight line best fitting it in the interval $L$~\cite{Mehta1990} and the power spectrum of the deviation of the $q$th nearest-neighbor spacing from its mean value $q$, $\eta_q=\epsilon_{q+1}-\epsilon_1-q$.

 The power spectrum is given in terms of the Fourier spectrum from 'time' $q$ to $k$, $S(k)=|\tilde{\eta}_k |^2$, with
\begin{equation}
\tilde{\eta}_k=\frac{1}{\sqrt{N}}\sum_{q=0}^{N-1} \eta_q\exp\left(-\frac{2\pi ikq}{N}\right)
\label{delta}
\end{equation}
when considering a sequence of $N$ levels.  It was shown in Refs.~\cite{Relano2002,Faleiro2004}, that for $\tilde k=k/N\ll 1$ the power spectrum  of level fluctuations  for complete level sequences exhibits a power law dependence $\langle S(\tilde k)\rangle\propto (\tilde k)^{-\alpha}$. Here, for regular systems $\alpha =2$ and for chaotic ones $\alpha =1$ regardless of whether time reversal symmetry is preserved or not. The power spectrum and the power law behavior were studied numerically in Refs.~\cite{Robnik2005,Salasnich2005,Santhanam2005,Relano2008} and experimentally in  microwave billiards in Refs.~\cite{Faleiro2006,Bialous2016}. It was also successfully applied to the measured molecular resonances in $^{166}$Er and $^{168}$Er~\cite{Mur2015}. In Refs. \cite{Bialous2016,Dietz2017} it was shown that the long-range spectral fluctuations provide exceptionally useful statistical measures when we have to deal with the missing levels.

\section{Missing level statistics}
 In experimental investigations~\cite{Liou1972,Zimmermann1988,Agvaanluvsan2003} the completeness of energy spectra is a rather rare situation. The problem of missing levels can be circumvented in open systems by using the scattering matrix formalism. The fluctuation properties of the scattering matrix elements provide sensitive measures for the chaoticity, e.g., in terms of their correlation functions~\cite{Dietz2009,Dietz2010}, Wigner reaction matrix or the enhancement factor~\cite{Lawniczak2010,Lawniczak2015,Sirko2016}.

For closed  or weakly open systems or when the full scattering matrix, including scattering amplitudes and phases, is not available, analytical expressions were derived for incomplete spectra based on RMT in Ref.~\cite{Bohigas2004}. The fraction of detected resonances is characterized by the parameter $\varphi$, where $0<\varphi \leq 1$. For such systems the nearest-neighbor spacing distribution  $p(s)$ is expressed in terms of the $(n+1)$st nearest-neighbor spacing distribution $P(n,\frac{s}{\varphi})$
\begin{equation}
p(s)= \sum_{n=0}^{\infty}(1-\varphi)^{n}P(n,\frac{s}{\varphi}).
\label{pdistr}
\end{equation}
 For the GOE systems the first term in Eq.~(\ref{pdistr}) is well approximated by
\begin{equation}
P(0,\frac{s}{\varphi}) = \frac{\pi}{2}\frac{s}{\varphi}\exp\left[-\frac{\pi}{4}\left(\frac{s}{\varphi}\right)^{2}\right].
\label{p0distr}
\end{equation}
For the complete sequences, $\varphi$ = 1, $P(0,s)$ reduces to the well known Wigner surmise formula for the NNSD.

The second term $P(1,\frac{s}{\varphi})$ is defined by the formula
\begin{equation}
P(1, \frac{s}{\varphi})=\frac{8}{3\pi^{3}}\left(\frac{4}{3}\right)^5\left(\frac{s}{\varphi}\right)^4\exp\left[-\frac{16}{9\pi}\left(\frac{s}{\varphi}\right)^2\right].
\label{p1distr}
\end{equation}
For the complete spectra, $\varphi$ = 1, $P(1,s)$ is given by the NNSD of the symplectic ensemble with $\left<s\right>=2$.

The higher spacing distributions $P(n, \frac{s}{\varphi})$ for $n=2,3,\ldots$ are well approximated by their Gaussian asymptotic forms, centered at $n+1$

\begin{equation}
P(n,\frac{s}{\varphi})= \frac{1}{\sqrt{2\pi V^2(n)}}\exp\left[-\frac{(\frac{s}{\varphi}-n-1)^2}{2V^2(n)}\right],
\label{pndistr}
\end{equation}

with the variances

\begin{equation}
V^2(n) \simeq \Sigma^2(L=n)-\frac{1}{6}.
\label{vvariance}
\end{equation}
The number variance $\Sigma^2(L)$  in the Eq. (\ref{vvariance}) is the variance of the number of levels contained in an interval of length $L$ ~\cite{Mehta1990}.

In the spectral analysis of the experimental data the integrated nearest-neighbor spacing distribution $I(s)$  also plays an important role.  It is often used to distinguish between the systems possessing or lacking time-reversal symmetry, e.g., described by the GOE and GUE ones, where the $I(s)$ sensitive dependence at small level separations $s$ is especially important

\begin{equation}
I(s)= \int^s_0p(s')ds'.
\label{ipdistr}
\end{equation}

The spectral rigidity of the spectrum $\Delta_3(L)$ is a measure of the long-range fluctuation properties in the spectra. In the presence of missing levels \cite{Bohigas2004} the spectral rigidity of the spectrum $\delta_3(L)$ may be expressed in terms of those for the complete spectra $\Delta_3\left(L\right)$,
\begin{equation}
\delta_3(L)=(1-\varphi)\frac{L}{15} + \varphi^2\Delta_3\left(\frac{L}{\varphi}\right).
\label{delta3}
\end{equation}

\begin{figure}[h!]
\includegraphics[width=0.8\linewidth]{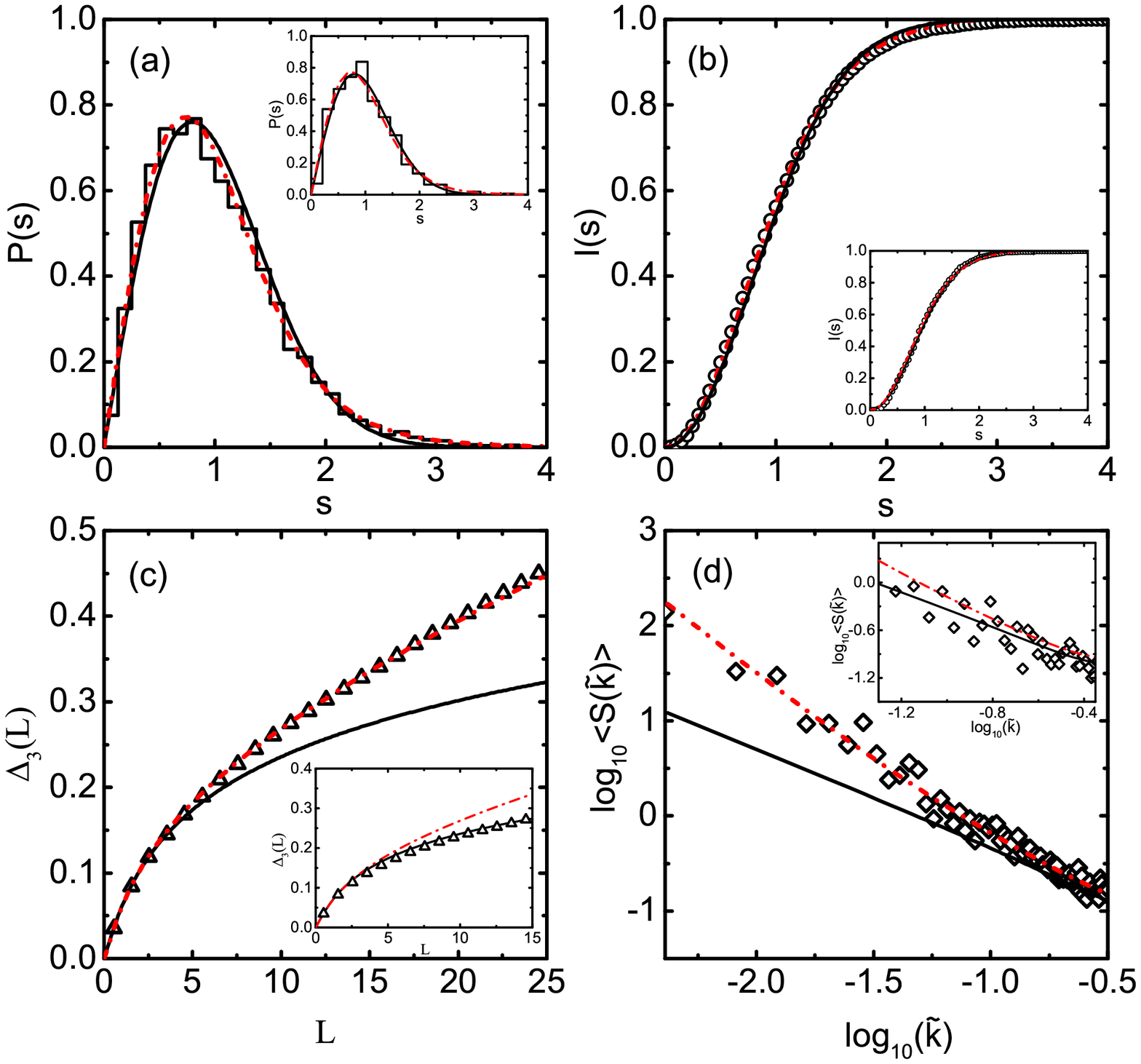}
\caption{
Spectral properties of the rescaled resonance frequencies of the 3D microwave cavity and the power spectrum of level fluctuations. Panels (a) and (b) show the nearest-neighbor spacing distribution (histogram) and the integrated nearest-neighbor spacing distribution  (circles).
The insets in panels (a) and (b) show the experimental results for the nearest-neighbor spacing distribution (histogram) and the integrated nearest spacing distribution (circles) obtained in the narrower frequency range 7-9 GHz. In these calculations the complete spectra ($\varphi=1$) for nine different realizations of the cavity were used (see the text).  The GOE predictions ($\varphi=1$) and the theoretical predictions evaluated for the fraction of observed levels $\varphi=0.89$ are shown in the insets with full and red broken lines, respectively.
The spectral rigidity of the spectrum (triangles) and the average power spectrum of level fluctuations (diamonds) are shown in panel (c) and (d),  respectively. The experimental results are compared to those of the eigenvalues of random matrices from the GOE (black full lines). The corresponding missing-level statistics (red broken lines) were calculated for the GOE system for $\varphi=0.89$.
The insets in panels (c) and (d) show the experimental results for the spectral rigidity (triangles) and the average power spectrum of level fluctuations (diamonds) obtained in the frequency range 7-9 GHz for the complete spectra ($\varphi=1$) for nine different realizations of the cavity. The GOE predictions ( $\varphi=1$) and the theoretical predictions evaluated for the fraction of observed levels $\varphi=0.89$ are shown in the insets with full and red broken lines, respectively.
}
\label{Fig3}
\end{figure}

\section{Power spectrum  of level fluctuations}

It is important to point out that in practice a priori knowledge about the expected fraction of observed levels $\varphi$ is not always possible. Therefore, in addition to the  spectral rigidity of the spectrum discussed above other sensitive tests of missing levels are of great value. It was demonstrated in Refs. \cite{Bialous2016,Dietz2017} that the power spectrum  of level fluctuations can be used as such a remarkably sensitive measure.  An analytical expression for the power spectrum  of level fluctuations in the case of incomplete spectra is given in Ref.~\cite{Molina2007},
\begin{eqnarray}
\langle s(\tilde k)\rangle &=&\nonumber
\frac{\varphi}{4\pi^2}\left[\frac{K\left(\varphi\tilde k\right)-1}{\tilde k^2}+\frac{K\left(\varphi\left(1-\tilde k\right)\right)-1}{(1-\tilde k)^2}\right]\\
&+& \frac{1}{4\sin^2(\pi\tilde k)} -\frac{\varphi^2}{12},
\label{noise}
\end{eqnarray}
which for $\varphi =1$ yields   the formula  for complete spectra $\langle S(\tilde k)\rangle$ ~\cite{Relano2002,Faleiro2004}.
Here, $0\leq \tilde k\leq 1$ and $K(\tau)$ is the spectral form factor, which equals $K(\tau ) = 2\tau -\tau \log(1+2\tau )$ for
the GOE systems.

In Fig.~3(a,b)  we show the experimental results obtained for the chaotic 3D microwave cavity for the nearest-neighbor spacing distribution (histogram) and the integrated nearest spacing distribution (circles), respectively.   Following  Ref.~\cite{Bohigas2004} the rescaled eigenvalues were normalized with average spacing unity. In the unfolding procedure we used a three-parameter fit $N(\nu)$  to the experimental data, discussed above. The GOE predictions ( $\varphi=1$) are marked in Fig.~3 by the full lines. The experimental results are in reasonable agreement with the GOE predictions showing that we are dealing with a system characterized by time reversal symmetry.  The theoretical predictions evaluated from equations ~(\ref{pdistr}) and ~(\ref{ipdistr}) are shown with red broken lines. They were evaluated for the fraction of the observed levels $\varphi=0.89 \pm 0.02$ which, on the basis of the spectral rigidity and the power spectrum  of level fluctuations analysis (panels (c) and (d) in Fig.~3, respectively), gave the best overall agreement with the experimental data. Ten terms in Eq.~(\ref{pdistr}) were used in the numerical calculations of the NNSD and the integrated nearest spacing distribution. A very good agreement of the experimental nearest-neighbor spacing distribution and the integrated nearest spacing distribution with the theoretical ones calculated for the observed levels $\varphi=0.89$ suggest that the observed departure from the GOE prediction is due to randomly missing levels.

This assumption was verified by the experimental results obtained in the narrower frequency range 7-9 GHz. In the insets in panels (a) and (b) we show the nearest-neighbor spacing distribution (histogram) and the integrated nearest spacing distribution (circles) obtained in this frequency range.  Analyzing  the measured spectra  and using in the analysis the fluctuating part of the number of levels we were able to identify, in the frequency range 7-9 GHz, the complete spectra ($\varphi=1$) for nine different realizations of the cavity. Such a procedure is described in details, e.g. in Ref. \cite{Gros2014}.  The GOE predictions ( $\varphi=1$) and the theoretical predictions evaluated for the fraction of observed levels $\varphi=0.89$ are shown in the insets with the full and red broken lines, respectively. The experimental results are in good agreement with the GOE predictions.

In Fig.~3 in the panels (c) and (d) we show the spectral rigidity of the spectrum  (triangles) and the average power spectrum of level fluctuations  (diamonds) for the 3D cavity. The theoretical predictions evaluated from the Eqs. ~(\ref{delta3}) and ~(\ref{noise})  for the fraction of the observed levels $\varphi=0.89$ are shown in Fig.~3(c) and Fig.~3(d), respectively, with  red broken lines. The fraction of the observed levels $\varphi=0.89 \pm 0.02$ was evaluated to get the best overall agreement with the experimentally found spectral rigidity and the power spectrum  of level fluctuations. Indeed,  Fig.~3(c) and Fig.~3(d) show that the experimental results are in good agreement with the numerical ones estimated on the basis of Eqs. ~(\ref{delta3}) and ~(\ref{noise}), respectively.  The GOE predictions are marked in Fig.~3(c) and Fig.~3(d) by full lines. Large departure of the experimental spectral rigidity of the spectrum and especially the average power spectrum  of level fluctuations  from the GOE predictions clearly demonstrate that the long-range spectral fluctuations provide powerful tools for the determination of the fraction of missing levels.
Moreover, our results suggest that these measures work well for the wave chaotic systems including the ones which do not simulate quantum systems.

For comparison, in the insets in the panels (c) and (d) we  demonstrate the experimental results for the spectral rigidity  (triangles) and the average power spectrum  of level fluctuations  (diamonds) obtained in the frequency range 7-9 GHz for the complete spectra ($\varphi=1$) for nine different realizations of the cavity. The GOE predictions ( $\varphi=1$) and the theoretical predictions evaluated for the fraction of observed levels $\varphi=0.89$ are shown in the insets with the full and red broken lines, respectively. In this case the experimental results are in agreement with the GOE predictions, proving that the departure from the GOE predictions observed in the panels (c) and (d) for $\varphi=0.89$  are solely due to randomly missing levels.

It is important to point out that finding all eigenfrequencies of the 3D microwave cavity in the frequency range 6-11 GHz was not the aim of this investigation. In the analysis of the experimental results we used 30 series of 215 identified eigenfrequencies which span this frequency range.  This kind of data is adequate for using the missing level statistics and to evaluate the fraction of observed levels $\varphi$.  The fraction of the observed levels $\varphi$ can also be independently estimated using some pieces of the complete spectra.    As describe above an analysis of  the measured spectra  allowed us to find the complete spectra in the narrower frequency range 7-9 GHz for nine different cavity configurations.   The fits of the experimental staircase functions to the formula $N_{W}(\nu)=A\nu^3-B\nu+C$ in the frequency range 7-9 GHz, where $A=\frac{8}{3}\frac{\pi}{c^3}V=(0.2259 \pm 0.0004)\times 10^{-27}$ s$^3$ is the volume coefficient in the Weyl formula (\ref{Bloch}), yielded the average coefficient $B=(1.442 \pm 0.174)\times 10^{-9}$ s and $C=-66.0 \pm 1.4$, respectively.  The constant $|C| \simeq 66$ is the cumulative number of levels of the 3D cavity in the frequency range 0 - 7 GHz. Using the fit results in the frequency range 6-11 GHz we found $\Delta N_W=245$ levels.

 The estimated from the missing level statistics analysis fraction of the observed levels $\varphi=0.89 \pm 0.02$ appeared to be in good agreement with the fraction $\frac{\Delta N_{exp}}{\Delta N_{W}} \simeq 0.88 $ calculated as a ratio of experimentally identified eigenfrequencies $\Delta N_{exp}=215$ in the frequency range $6-11$ GHz to the predicted ones $\Delta N_{W}=245$.

As discussed above, for different realizations of the 3D cavity in the frequency range $6-11$ GHz  we found close to $88\%$ of levels. For comparison, in the narrower frequency ranges $6-8.5$ GHz and $8.5-11$ GHz, for different cavity realizations, $89\%$ and $87\%$ of levels, respectively, were on average identified. Fig.~2 clearly demonstrates that for the individual realizations of the cavity, even for lower frequencies, the scattering matrix elements $|S_{11}|$ and $|S_{22}|$ show a strong dependence of the recorded resonances on the antennas positions. As a consequence of this behavior some resonances, often being well separated from the others and therefore not correlated, are not recorded at all. A good agreement of the experimental results with the missing level statistics strongly suggests that such resonances are randomly missed.

 The measured experimental spectra of the 3D cavity allowed us to test further the short- and long-range spectral fluctuations, also in the case of $\varphi < 0.89$. To do this we have chosen  two completely different strategies in removing the eigenvalues. In the first one we randomly removed the eigenvalues from the experimental spectra to get much lower fraction of the observed levels $\varphi=0.5$.
 In the panels (a-d)  in Fig.~4 we show the results obtained for the modified experimental spectra of the chaotic 3D microwave cavity with  $\varphi=0.5$ for the nearest-neighbor spacing distribution (histogram), the integrated nearest spacing distribution (circles), the spectral rigidity of the spectrum (triangles), and the power spectrum  of level fluctuations (diamonds), respectively. The experimental results are compared with the theoretical predictions (red broken lines) calculated from Eqs. ~(\ref{pdistr}), ~(\ref{ipdistr}),  ~(\ref{delta3}), and  ~(\ref{noise}), respectively. For a comparison, the GOE predictions  are shown in full lines.  Fig.~4 clearly demonstrates that even for the lower fraction of the observed levels $\varphi=0.5$ an agreement between the experimental data and the theoretical ones is very good. This reconfirms that the missing level statistics and the power spectrum  of level fluctuations are powerful tools which can be also successfully applicable for the spectra analysis of wave chaotic systems such as 3D microwave cavities.

\begin{figure}[h!]
\includegraphics[width=1.0\linewidth]{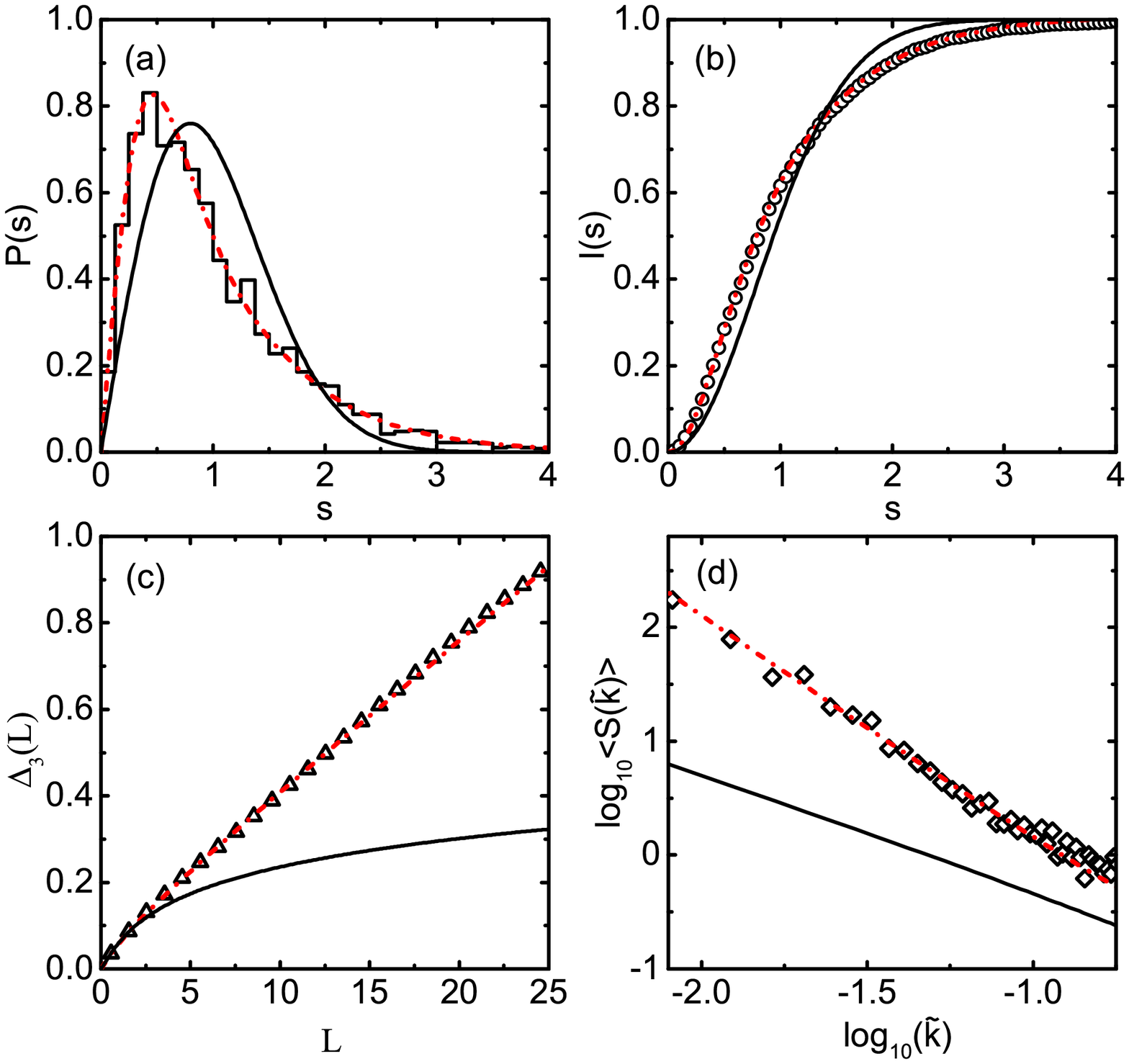}
\caption{
Spectral properties of the rescaled resonance frequencies of the 3D microwave cavity and the power spectrum of level fluctuations calculated for the modified spectra with randomly removed resonances with the fraction of the observed levels $\varphi=0.5$.   Panels (a-d) show the nearest-neighbor spacing distribution  (histogram), the integrated nearest-neighbor spacing distribution  (circles), the spectral rigidity of the spectrum  (triangles), and the average power spectrum of level fluctuations (diamonds), respectively. The experimental results are compared to those of the eigenvalues of random matrices from the GOE (full lines). The corresponding missing-level statistics (red broken lines) were calculated for the GOE system for $\varphi=0.5$.
}
\label{Fig4}
\end{figure}

In the second approach we applied the following procedure: in every experimental spectrum with $\varphi = 0.89$ we identified the doublet of the closest eigenvalues and removed one of them. The procedure was repetitively applied until we obtained the spectra with $\varphi = 0.7$ where additionally to randomly missing states the clusters of the closest levels were treated as the overlapping ones. In the panels (a-d)  in Fig.~5 we show the results obtained for the modified experimental spectra of the chaotic 3D microwave cavity with  $\varphi=0.7$ for the nearest-neighbor spacing distribution (histogram), the integrated nearest spacing distribution (circles), the spectral rigidity of the spectrum (triangles), and the power spectrum  of level fluctuations (diamonds), respectively. The experimental results are compared with the theoretical predictions for the randomly missing levels (red broken lines) calculated from  Eqs. ~(\ref{pdistr}), ~(\ref{ipdistr}),  ~(\ref{delta3}), and  ~(\ref{noise}), respectively. For a comparison, the GOE predictions  are shown by  full lines.
Fig.~5 demonstrates that the level statistics in the experimental incomplete spectra in which we deal with both randomly missing levels and the unresolved ones cannot be properly described by the standard missing level statistics. In order to properly describe the experimental short- and long-range level missing statistics in such a case, one should apply the RMT calculations. In the calculations we generated ensembles of 99 random matrices with the dimension $M=295$. In the unfolding procedure we used the polynomial of the fifth order. For further calculations we used the 99 sets of 245 eigenvalues which remained after removing the first and the last 25 eigenvalues of the matrices. The number of the remaining eigenvalues of the matrices was chosen to be equal to the number of eigenfrequencies $\Delta N_W=245$ of the cavity in the frequency range 6-11 GHz.  Furthermore, from each set of 245 eigenvalues, similarly to the situation with the experimental data, we first randomly removed 11\% of states and then applied the procedure of removing clustering states to get  $\varphi = 0.7$. The RMT results are shown in panels (a-d) with the green dotted lines. The agreement between the experimental data and the RMT ones is good. This shows that even in the case of strongly overlapping states the RMT analysis can be useful for identifying of the fraction of the observed levels.

\begin{figure}[h!]
\includegraphics[width=1.0\linewidth]{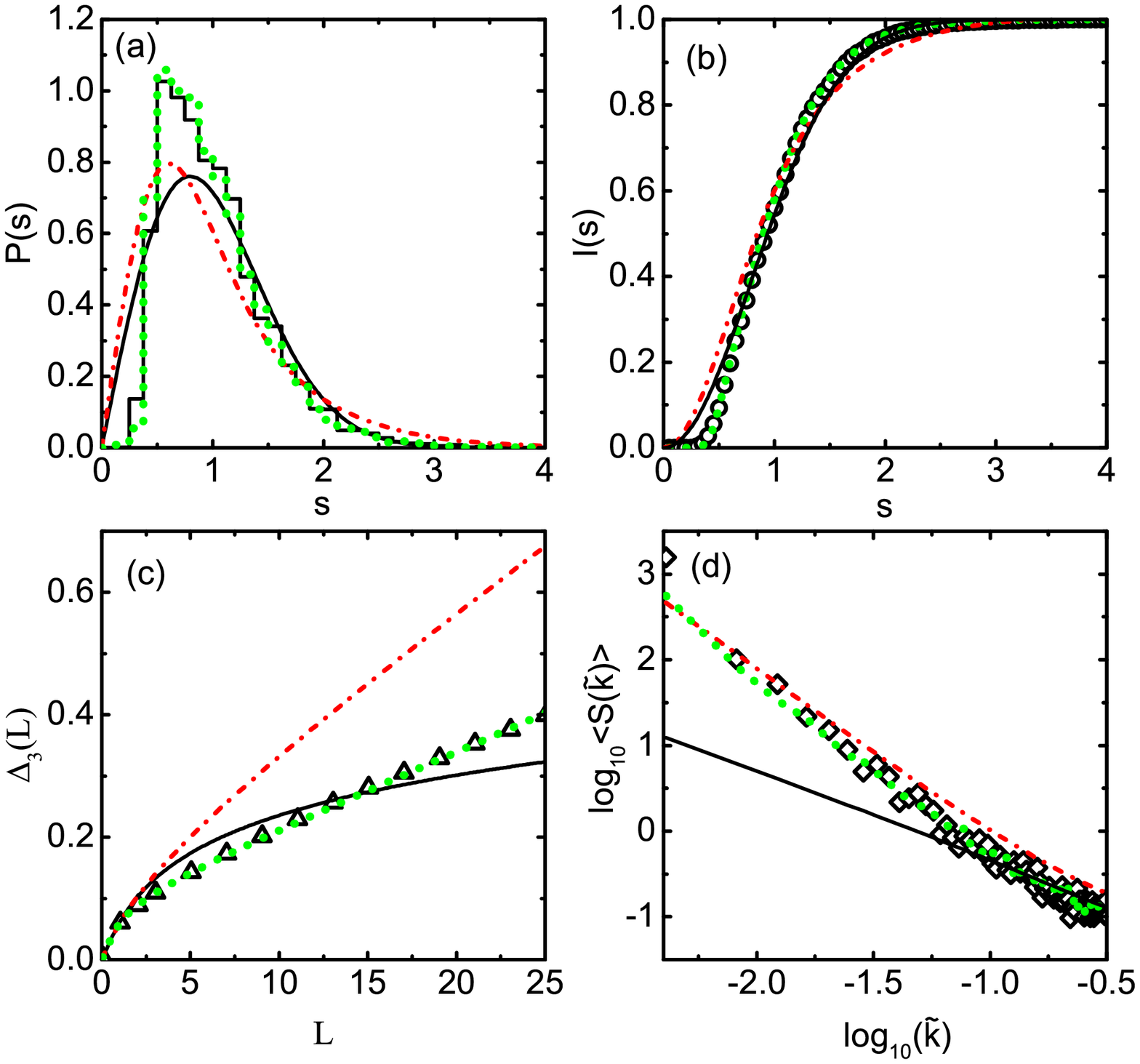}
\caption{
Spectral properties of the rescaled eigenvalues of the 3D microwave cavity and the power spectrum of level fluctuations calculated for the modified experimental spectra from which the closest resonances were removed to get the fraction of the observed levels $\varphi=0.7$.   Panels (a-d) show the nearest-neighbor spacing distribution  (histogram), the integrated nearest-neighbor spacing distribution (circles), the spectral rigidity of the spectrum  (triangles), and the average power spectrum of level fluctuations (diamonds), respectively. The experimental results are compared to those of the eigenvalues of random matrices from the GOE (full lines). The corresponding missing-level statistics for randomly removed states (red broken lines) were calculated for the GOE system for $\varphi=0.7$. The RMT results shown with the green dotted lines are also attained for the incomplete spectra with $\varphi = 0.7$ but in this case the spectra were obtained by randomly removing of 11\% states and then applying the procedure of removing clustering states.
}
\label{Fig5}
\end{figure}

\section{Conclusions}

We present the first experimental study of the fluctuation properties in incomplete spectra with randomly missing levels of the microwave chaotic 3D cavity. The experimental results are in good agreement with the analytical expressions for missing level statistics, Eqs.~(\ref{pdistr})-(\ref{delta3}),   and for the power spectrum of level fluctuations,  Eq.~(\ref{noise}). All these expressions explicitly take into account the fraction of observed levels $\varphi$. The spectral rigidity of the spectrum, and particularly the power spectrum  of level fluctuations, appeared to be very sensitive to it. Therefore, we used them to determine the fraction of observed levels, $\varphi=0.89\pm 0.02$, in the experimental spectra.  This fraction is in good agreement with the fraction $\varphi=0.88$ calculated as a ratio of experimentally identified eigenfrequencies to the predicted ones from the Weyl formula. The excellent agreement between the experimental and analytical results, clearly shown in Fig.~\ref{Fig3} and Fig.~\ref{Fig4}, proves the potency of the  missing level statistics and the power spectrum  of level fluctuations for the description of wave-chaotic systems, such as irregular microwave 3D cavities. In such systems either due to too small amplitudes of some resonances or the cubic dependence of the number of levels on the frequency  $\nu$, which for larger $\nu$ may cause overlapping of some resonances, the randomly missing levels in the measurements are unavoidable. We also show (see Fig.~\ref{Fig5}) that in the case of incomplete spectra with many unresolved states, when lost states are mainly due to their overlap, the above-mentioned procedures may fail. In such a case the RMT calculations can be useful for the determination of the fraction of observed levels.

This work was supported in part by the Ministry of Science and Higher Education grants UMO-2016/23/B/ST2/03979 (LS) and UMO-2013/09/D/ST2/03727 (M{\L}). The authors sincerely thank Alexander Wlodawer for careful reading of the manuscript.

\end{document}